\documentclass[pra,aps,twocolumn,showpacs,epsf]{revtex4}
\usepackage{graphicx}
\usepackage{epsfig}
\usepackage{epsf}
\usepackage{amssymb}
\usepackage{epstopdf}
\usepackage{amsmath}
\usepackage{amstext}
\usepackage{latexsym}
\usepackage[matrix,frame,arrow]{xypic}

\begin{document}

\title{Efficient construction of 2-D cluster states with probabilistic quantum gates}
\author{ Qing Chen$^{1,2}$}\email{chenqing@mail.ustc.edu.cn}
\author {Jianhua Cheng$^2$}
\author{ Ke-Lin Wang$^2$}
\author {Jiangfeng Du$^{2,3,4}$}\email{djf@ustc.edu.cn}

\address{$^{1}$Key Laboratory of Quantum Information, University of Science and Technology of China, CAS, Hefei 230026, PR China\\
$^{2}$Department of Modern Physics, University of Science and Technology of China, Hefei 230026, PR China\\
$^{3}$Hefei National Laboratory for Physical Sciences at Microscale,
University of Science and Technology of China, Hefei, Anhui 230026, PR China\\
$^{4}$Department of Physics, National University of Singapore, 2 Science Drive 3, Singapore 117542
}
\date{\today}
\begin{abstract}
We propose an efficient scheme for constructing arbitrary 2-D cluster states using probabilistic entangling quantum gates.
In our scheme, the 2-D cluster state is constructed with star-like basic units generated from 1-D cluster chains.
By applying parallel operations, the process of generating 2-D (or higher dimensional) cluster states is significantly accelerated, which
provides an efficient way to implement realistic one way quantum computers.

\end{abstract}
\maketitle

Quantum computation offers a potentially exponential computational
speed-up over classical computers in certain tasks, which makes it one of the most promising ways
 for the future's computations \cite{1}.
Nowadays conceptual quantum computer with few qubits has been
demonstrated experimentally in many systems \cite{2}; however, how
to extend these systems to a large number of qubits, known as
scalability still remains a question. While in most of the current
experiments, quantum logic gates are implemented with sequences of
highly controlled interactions between selected particles, Robert
Raussendorf and Hans J. Briegel \cite {3,4} recently proposed a
different model of a scalable quantum computer, namely the one way
quantum computer, which constructs quantum logic gates by
performing single qubit measurements on cluster states. In this
model, the total resource needed for the quantum computation is a
2-D (two dimensional) cluster state prepared at numerous qubits
(e.g. a lattice-like cluster), which serves as a ``substrate" for
the computation. After the preparation of this universal
``substrate", the remaining work is to perform single qubit
measurements and,
 the final results are read out from those qubits that were not
measured in the whole process.

Optical approaches to quantum computation are attractive because
of the long decoherence time of photons and near perfect single
qubit operations. However, due to the lack of interactions between
photons, the main challenge within these approaches is the
realization of two-qubit logic gates. A breakthrough is the Knill,
Laflamme and Milburn's (KLM) scheme \cite{5}, in which an
entangling gate was proposed based on linear optics. Nevertheless,
to implement a near-deterministic gate within this scheme requires a large number of optical elements,
which makes the present scalable quantum computation inefficient.
Moreover, with the restriction of involving only static linear optical systems, there are limitations for the success
probability of such kind of entangling gates \cite {Scheel,Eisert}.
In recent works [8-15,23], the one way quantum computer scheme,
carried out in a cluster state, was introduced to implement efficient scalable
quantum computations with probabilistic entangling gates.
Specifically, Ref. \cite{Daniel,11}
show efficient schemes to construct 1-D cluster states and suggest that 2-D cluster states be built from ``T-pieces".
Later, L.-M. Duan \emph{et al.} \cite{12}
presented an efficient scheme to construct 2-D lattice-like
cluster states. To achieve this goal, they made use of a ``cross"
(``+" shape) cluster state as the basic unit, which is built from
two long 1-D chain cluster states.
 In their proposal, the temporal overheads increase logarithmically with the qubit number $N$ and polynomially with $1/p$.
This result opens up the promising prospect to realize efficient quantum computation with probabilistic entangling gates.

In this paper, we report that an arbitrary 2-D cluster state can be directly constructed from star-like cluster states (bottom right in Fig.1),
which can be obtained by
performing single qubit measurements on a chain-like cluster state (top in Fig.1).
Moreover, in this scheme,  the time consuming of generating a 2-D cluster state is nearly equal to that
of generating a 1-D chain cluster state.

Specifically, following Duan's assumptions \cite{12},
one can reliably perform
two-qubit controlled phase flip (CPF) gates with a small success
probability $p$.  Also, all single qubit operations are regarded as perfect, which
are well justified in linear optical
systems. And in the calculation of temporal overheads, compared with the operation time of the CPF gate, we
neglect the time consuming of all single-qubit operations
including the single-qubit measurements.

Before stating our proposal, we list some important properties of
cluster states for future use \cite {16}. Since cluster state is relevant to the graph theory, we start with
graph language. For a given simple undirected graph
$\textsl{G}=\textsl{(V, E)}$, where $V$ denotes the vertex set and
$E(A,B)=\{\{a,b\}\in E:a\in A,b\in B,a\neq b\}$ denotes the set of
edges between the vertices, one can define a set of commutable
operators
$\textsl{K}^i=\sigma_{x}^i\prod_{j\in N_i}\sigma_{z}^j$ regarding
every vertex of the graph, here $i$ stands for an arbitrary
vertex of the graph and $N_i$ refers to the set of vertex $i$'s
neighboring vertices, $\sigma_{x,y,z}^i$ denote the
Pauli matrices corresponding to the $i$-th vertex. The cluster state
is then defined as the co-eigenstates of the group of operators.
The local Pauli measurements on cluster states induce some interesting results. Specifically,
$\sigma_{z}^i$
measurement deletes all edges incident to vertex $i$, while $\sigma_{y}^i$ measurement simply replaces the subgraph
$G[N_i]$ by its complement $G[N_i]^c$.
In the case of the $\sigma_x$ measurement, the resulting state is a cluster state associated with
the graph $$\textsl{G}^{\;\prime}=\textsl{G}\vartriangle\textsl{E}(N_j,
N_i)\vartriangle\textsl{E}(N_j\cap N_i,N_j\cap
N_i)\vartriangle\textsl{E}(\{j\},N_i-\{j\}),$$ where
$\textsl{E}\vartriangle\textsl{F}=(\textsl{E}\cup\textsl{F})-(\textsl{E}\cap\textsl{F}),$
and $j\in N_i$ \cite {note2}.

Next we will show that once the star-like cluster state (depicted in bottom right of Fig. 1) is prepared, one can construct an arbitrary 2-D cluster
state directly, and the time consuming of this process is the single CPF gate operation time $t_a$.
For example, an $N$ qubits lattice cluster state can be generated with $N$ such basic
units, each center qubit corresponding to one site of the lattice.
Now we  connect these basic units to a 2-D lattice-like cluster state via probabilistic CPF gates.
Suppose each unit has $n_l$ ``arms", there will be $n_l/4$ connections between each adjacent pair of sites, and
all these operations are to be done at the same time
(Note here parallel operations are introduced in this process).
At least one successful connection is needed between each pair of sites.
After the connections are completed, we keep one successful connection and
remove the others by $\sigma_z$ measurements on
the qubits adjacent to the center qubits, no matter they succeed or not.
Next, we trim the cluster state by applying $\sigma_y$ measurements on the four
redundant qubits of the only one kept successful connection, thus the required connections between center qubits are established and a 2-D lattice-like
cluster is successfully constructed.
Now we should find a proper $n_l$ so that the lattice
cluster state can be created at a satisfying probability. Let $p_c$ be the probability of at least one
connection succeeds between a pair of sites.
And one can easily get $p_c=1-(1-p)^{n_l/4}$. Consequently, the total success probability of the lattice cluster
state is $P_a={p_c}^{2N}$.  Naturally we order
$P_a>1-\epsilon$, where $\epsilon$ is a small positive number standing for the overall failure probability.
If $N$ is sufficiently large and $p$ is small, we get $n_l\simeq(4/p)\ln(2N/\epsilon)$.

\begin{figure}
\begin{center}
\epsfig{file=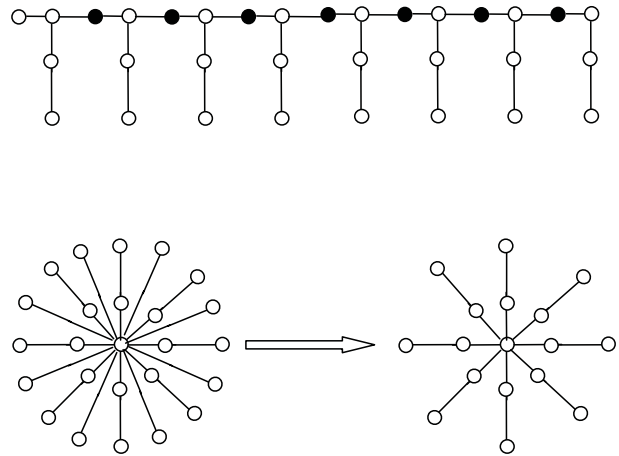,width=0.45\textwidth,height=6cm}

\end{center}
\caption{
Illustration of generation process of the required star-like basic unit (bottom right) from 1-D cluster chain (top).
Performing $\sigma_x$ measurements on qubits without ``arms" (black circles on the top figure) to reach the bottom left, followed by performing
$\sigma_z$ measurements on the one-qubit ``arms" (bottom left), we finally obtain cluster states on the bottom right. }
\end{figure}

 By now we showed that it is convenient to generate a 2-D
cluster state with such basic units.
Here we give a method to create these basic units by performing $\sigma_x$ and $\sigma_z$ measurements on chosen qubits of a certain cluster
state (see top in Fig. 1).  Given
a $2n_l$ qubits (main chain length) cluster state with half of them having a two-qubit ``arm", by performing $\sigma_x$ measurements on those
qubits which do not have an ``arm",
one achieves the expected star-like basic unit with $n_l$ ``arms"(In fact, there exists $n_l$ one-qubit
``arms", which can be easily removed by performing $\sigma_z$ measurements on them).
Note that all operations involved in the process are single qubit measurements and local unitary transformations.

\begin{figure}[t]
\begin{center}
\epsfig{file=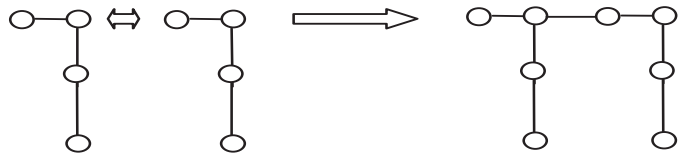,width=0.5\textwidth,height=2cm}
\end{center}
\caption{
The figure of generating extended cluster chain with four-qubit
cluster chains.}
\end{figure}

In a slightly different way from Duan  $\emph{et al}$'s proposal \cite{12}, the 1-D chain cluster state
with "arms" can be generated efficiently. The detailed process is shown below.

Suppose we have some $n_0$ qubits cluster chains with half of
the qubits having a two-qubit ``arm"($n_0$ stands for the number of
qubits in the  main chain and is supposed to be large enough), we connect two of
them together with a probabilistic CPF gate. If this attempt fails,
by applying a $\sigma_z$ measurement, two
qubits at the end of each cluster's main chain are removed. And in the
next round we connect the left two
$(n_0-2)$ cluster chains. The expected length of the connected cluster chain can be calculated as $n_{1}=%
\sum_{i=0}^{n_{0}/2}2\left( n_{0}-2i\right) p\left( 1-p\right)
^{i}\approx 2n_{0}-4\left( 1-p\right) /p$. So if we want to get an expected longer
cluster chain after a successful connection, $n_0$ should be
larger than a critical length $n_c=4(1-p)/p$. Iterating this
process, one can get an arbitrary length cluster chain. After $r$
rounds successful connection, the total chain length $n_r$, the
time consuming $T_r$, and the total number of
attempt $M_r$ follow the recursion rules:

\begin{eqnarray}
n_{r}&=&2n_{r-1}-n_{c}, \nonumber \\
T_{r}&=&T_{r-1}+t_{a}/p, \\
 M_{r}&=&2M_{r-1}+1/p. \nonumber
\end{eqnarray}

In writing the recursion rule for $T_r$, one should keep in mind that two cluster chains for each connection are assumed to be
prepared in parallel.
From the above three recursion rules, one gets $n_r=(n_0-n_c)2^r+n_c, T_r=T_0+rt_a/p$, and $M_r=(M_0+1/p)2^r-1/p$.
According to the first formula, $r$ can be expressed with the main chain length $n$ ($n_r$) after $r$th rounds successful
connections, which can be written as $r=\log_2[(n-n_c)/(n_0-n_c)]$. Replacing $r$ in the left two formulae with this
equation, one finds that the total time consuming $T (n)$ and the number of attempts $M (n)$ scales with the main chain length
$n$ as

\begin{equation}
\label{Tn}
T\left( n\right) =T_{0}+\left(
t_{a}/p\right) \log _{2}\left[ \left( n-n_{c}\right) /\left(
n_{0}-n_{c}\right) \right],
\end{equation}
\begin{equation}
\label{Mn}
M\left( n\right) =\left(
M_{0}+1/p\right) \left( n-n_{c}\right) /\left( n_{0}-n_{c}\right)
-1/p ,
\end{equation}

respectively, where $T_0$ and $M_0$ stand for the time and attempts needed for a $n_0$ qubits chain cluster state.

We now figure out the preparation of an $n<n_c$ qubits
cluster chain with the required ``arms". Started from single
qubits, we first create four-qubit cluster chains,
then connect two of them in the
way illustrated in Fig.2. As a result, a four-qubit cluster chain with two two-qubit ``arms" is constructed. By iterating this process,
one can reach an arbitrary length
such cluster chain.
Generally speaking, given two $2^i(i>1)$ qubits
cluster chains, by applying a probabilistic CPF gate on them successfully, we can get a $2^{i+1}$ qubits cluster chain.
If this attempt fails, we go back to the single qubit and try the whole process again. Let $T_i$ and $M_i$ be
the preparation time and the number of attempts respectively, they obey the following recursion rules $T_{i}=(1/p)\left(
T_{i-1}+t_{a}\right) $ and $M_{i}=(1/p)\left( 2M_{i-1}+1\right)$. With $T_1=t_a(1/p^2+1/p)$ and $M_1=2/p^2+1/p$ (for
preparing a four-qubit chain cluster),
one gets the scaling rules
$T\left( n\right) \simeq t_{a}\left( 1/p\right) ^{\log _{2}n+1}$ and $%
M\left( n\right) \simeq \left( 2/p\right) ^{\log _{2}n+1}/2$.

To create an $n>n_c$ cluster chain, one needs to combine the above two protocols. Let $n_0=n_c+1$, then
$T_0=t_a(1/p)^{\log_2(n_c+1)+1}$ and $M_0=(2/p)^{\log_2(n_c+1)+1}/2$. Replacing these equations to
Eq. (\ref{Tn}) and (\ref{Mn}),
the overall time consuming
$T(n)$ and the number of attempts $M(n)$ are:

\begin{equation}
\label{T-n}
T\left( n\right) \simeq t_{a}\left( 1/p\right) ^{\log _{2}\left(
n_{c}+1\right)+1 }+\left( t_{a}/p\right) \log _{2}\left( n-n_{c}\right),
\end{equation}
\begin{equation}
\label{M-n}
M\left( n\right) \simeq \left( 2/p\right) ^{\log _{2}\left( n_{c}+1\right)+1
}\left( n-n_{c}\right) /2.
\end{equation}

\begin{figure}[t]
\epsfig{file=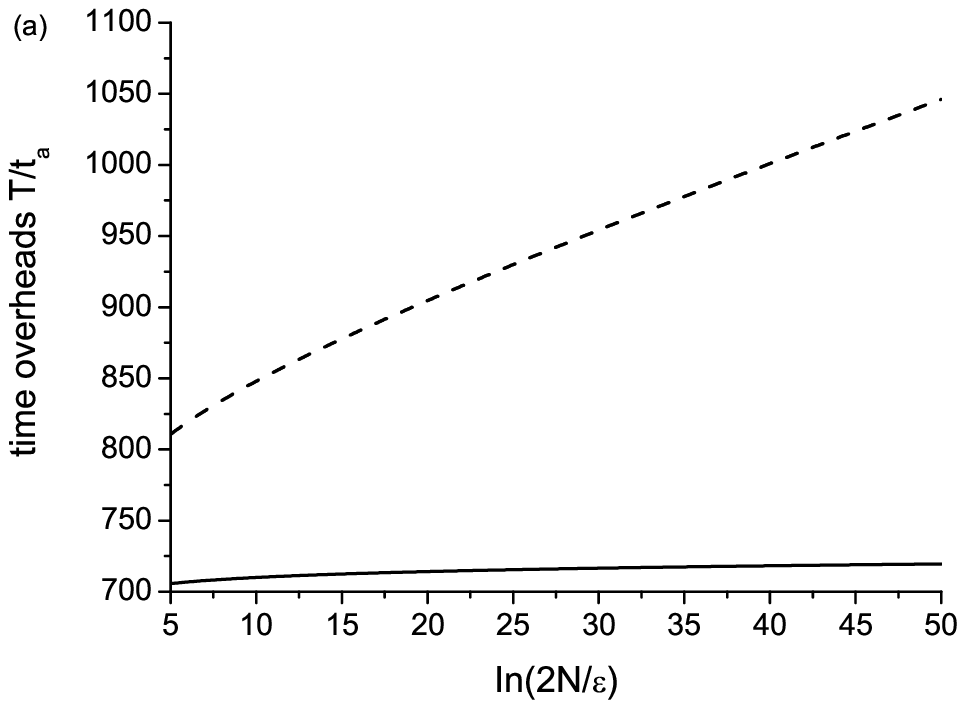, width=6cm, height=5cm}
\epsfig{file=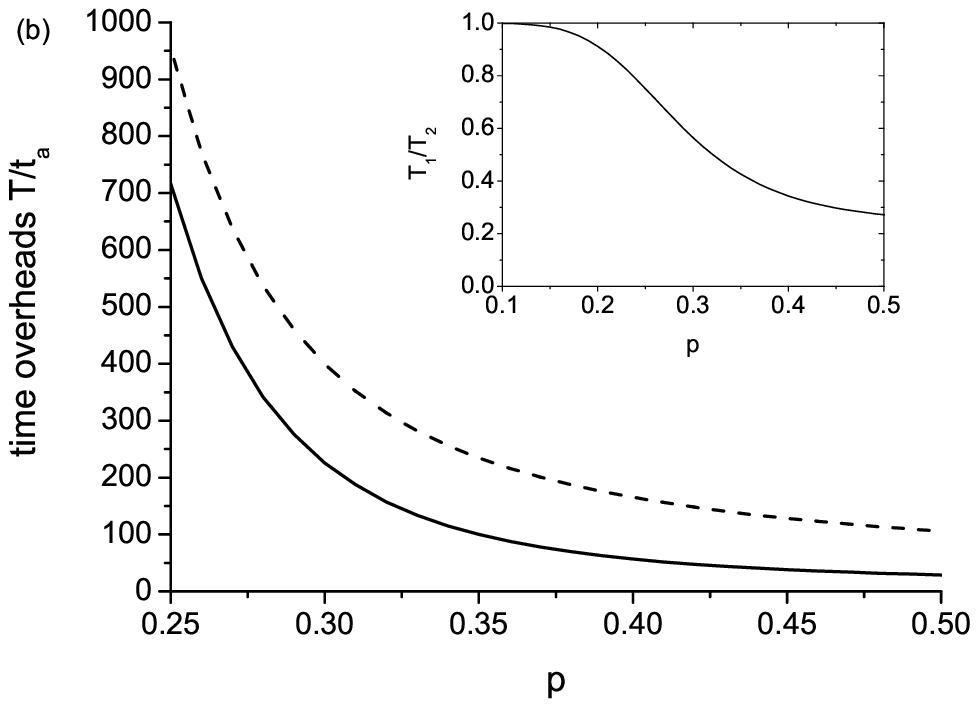, width=6cm, height=5cm}
\caption{
 $T-p$, $T-\ln(2N/\epsilon)$ relations. $T_1$ (solid line) stands for time overheads in
our scheme, while $T_2$ (dash line) as in Duan $\emph{et al}$'s scheme \cite{12,note}. In
Fig. 3a, we set $p=0.25$. In Fig. 3b, parameter $\ln(2N/\epsilon)$
is chosen as $30$. The inserted figure in Fig. 3b plots dependence
of $T_1/T_2$ on $p$. \cite {note3}}
\end{figure}

Taking the construction
of a 2-D $N$ qubits lattice-like cluster state as an example, we now calculate the time overheads and the number of
attempt needed for constructing a 2-D cluster state. As stated above,
if we want to ensure the total success probability of the construction
to be larger than $1-\epsilon$, the basic unit should have $n_l=(4/p)\ln(2N/\epsilon)$ ``arms".
To construct a square lattice cluster state of $N$ qubits, we need $N$ such basic units in total, and
each basic unit is made from a 1-D $2n_l$ qubits chain cluster with half of the qubits in the chain
having a two-qubit ``arm".
Consequently, we need to prepare $N$ such chain
clusters (this can be done in parallel), which requires $NM(2n_l)$ CPF attempts within a time period of $T(2n_l)$ (refer to Eqs. (\ref{T-n})
and (\ref{M-n}) for expressions of $M(n)$ and $T(n)$). The final step is to connect the $N$ basic units to form an $N$ qubits
square lattice cluster state. Note that all the connections can be done in parallel in this process, so the whole
connection takes on average $(N/2)n_l$ CPF attempts in a single CPF operation time $t_a$. Summarizing these results,
one gets the the overall time consuming
and the total number of attempts as follows:

\begin{equation}
\label{finalT}
\begin{array}{rcl}
T\left( N\right)  & \simeq  & t_{a}\left( 1/p\right) ^{\log _{2}\left(
4/p-3\right)+1 } \\
& + & \frac{t_{a}}{p}\log _{2}\left( \frac{4}{p}\left[ 2\ln \left(
2N/\epsilon \right) -1\right] \right)  \\
& + &t_a, \\
\end{array}
\end{equation}
and
\begin{equation}
\label{finalM}
\begin{array}{rcl}
M\left( N\right) &\simeq& \left( 2/p\right) ^{2+\log _{2}\left( 4/p-3\right) }N
\left[ 2\ln \left( 2N/\epsilon \right) -1\right] \\
& + & \frac{2N}{p}\ln(2N/\epsilon).\\
\end{array}
\end{equation}

We now take a close look at what we have achieved. Since
$2N/\epsilon$ is not likely larger than $e^{50}$, it is reasonable
that we set $\ln(2N/\epsilon)$ between $5$ and $50$. Fig. 3
compares the overall time overheads between Duan \emph{et al}'s
scheme ($T_2$) \cite{note} and ours ($T_1$) that is needed for
constructing a 2-D lattice-like cluster state. Fig. 3(a) shows
that the overall time overheads in our scheme scales
$\ln(2N/\epsilon)$ from linear to logarithmic compared with Duan
\emph{et al}'s proposal. From Fig. 3(b), we see that $T_1$ is
always smaller than $T_2$ as $p$ varies. More interestingly, the
inserted $T_1/T_2-p$ relation in Fig. 3(b) indicates that our
scheme have a relatively more significant improvement when $p$ is
larger than $0.2$.
At last, note that to minimize the time overheads, we implement parallel
operations during the whole process. As a result, it
costs more attempts within our scheme.
Nevertheless, the numbers of attempts in the two
schemes have the same scaling with qubit number $N$, specifically,
$N\ln(2N/\epsilon)$.

\begin{figure}[t]
\begin{center}
\epsfig{file=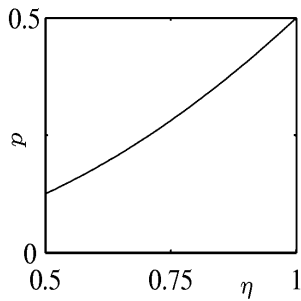, width=4cm, height=4cm}
\end{center}
\caption{
Success probability $p$ versus the detector efficiency $\eta$ with no spontaneous emission. The Jaynes-Cummings coupling
coefficient $g=0.3$, the leacage rate of the cavity $\kappa=1$
\cite {11}.
}
\end{figure}

The above results can be analyzed qualitatively from the Eq. (\ref{finalT}) and (\ref{duanT}). Note that our improvement
lies in the process of constructing 2-D cluster states,
which is reflected by substituting the third term $\frac{t_{a}}{p}\ln \left( 2N/\epsilon \right)$ in Eq. (\ref{duanT})
with $t_a$ in Eq. (\ref{finalT}). For example, if $\ln(2N/\epsilon)=13$,
the first term is comparable with $\frac{t_{a}}{p}\ln \left( 2N/\epsilon \right)$ when $p$ is around $0.4$.
However,  as $p$ decreases, the first term increases much more rapidly than $\frac{t_{a}}{p}\ln \left( 2N/\epsilon \right)$. And
when $p$ is very small, the first term dominates the total time overheads and the other two terms are negligible.

Moreover, with our basic units, one can directly create an arbitrary 2-D
cluster state, which needs not to be the lattice like cluster state.
Fig. 5 shows a sketch map of a hexagonal cluster state generated
with our basic unit. Following the same calculating process,
one gets the time overheads $T(N)$ and the number of $M(N)$ 's relation with parameters $N$ (qubit number), $p$ ,
and $\epsilon$ , as stated below:

\begin{equation}
\begin{array}{rcl}
T\left( N\right)  & \simeq  & t_{a}\left( 1/p\right) ^{\log _{2}\left(
4/p-3\right)+1 } \\
& + & \frac{t_{a}}{p}\log _{2}\left( \frac{4}{p}\left[ \frac{3}{2}\ln \left(
3N/{2\epsilon }\right) -1\right] \right)  \\
& + &t_a, \\
\end{array}
\end{equation}
and
\begin{equation}
\begin{array}{rcl}
M\left( N\right) &\simeq& \left( 2/p\right) ^{2+\log _{2}\left( 4/p-3\right) }N
\left[ \frac{3}{2}\ln \left( 3N/{2\epsilon} \right) -1\right] \\
& + & \frac{3N}{2p}\ln(3N/{2\epsilon)}.\\
\end{array}
\end{equation}
Also,
without making any variation, these basic units can be used to efficiently build
a cluster state in any dimension.

\begin{figure}
\epsfig{file=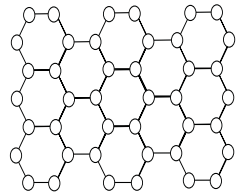, width=3cm, height=2.5cm}
\caption{Illustration of a 2-D hexagonal cluster state. }
\end{figure}

The prospect of implementing quantum computation with probabilistic entangling operations is relevant for a number of
experimental systems involving atoms, ions and photons \cite{5,duan1,Kuzmich}. In particular,
Barrett and Kok (BK) proposed a practical probabilistic entangling gate using spatially
separated matter qubits and single photon interference \cite {11}. They also demonstrated that this probabilistic entangling
gate is sufficient to construct cluster states. More interestingly, our scheme has the best efficiency when the success
probability of CPF gate is between $0.2$ and $0.5$ (See Fig. [3]). Note that in BK scheme, the success probability $p$
varies from $0.1$ to $0.5$ as the detector efficiency goes from $0.5$ to $1$ (See Fig. [4]), which makes BK scheme an
excellent physical implementation of our proposal. Also, we notice that the recently proposed
``repeat-until-success" scheme can eventually lead to deterministic gates \cite {Kwek,Lim}, which sheds light on
 direct implementation of any scalable quantum logic circuit.
However, in realistic systems, where photon loss can not be ignored, this scheme can not produce
entangling gates with unitary success probability.
In this case, the one way computer scheme appears to be an effective alternative to implement scalable quantum
computation, where our scheme proves to be useful.

To sum up, we have shown that using probabilistic CPF gates, one
can construct an arbitrary 2-D (or higher dimensional) cluster
state with the star-like basic units. Through single qubit
measurements, these basic units can be generated from a 1-D
cluster chain with two-qubit ``arms". The process scales
efficiently with the qubit number and the inverse of the success
probability. More interestingly, in our proposal, by applying
parallel operations in the process of building 2-D cluster states,
the time needed for constructing a 2-D and 1-D cluster state have
the same order of magnitude. This result is helpful to construct
2-D cluster states with arbitrary size, which is essential to the
implementation of experimental one way quantum computers.

We have benefited from Prof. Lu-Ming Duan's recent lecture at \emph{USTC}.
We thank Dr. Mian-Lai Zhou and Mr. Chenyong Ju for valuable
discussions and careful reading of the manuscript. We also thank Daniel E. Browne and Pieter Kok for
critical remarks.
This work is supported by NUS Research Project (Grant No.
R-144-000-071-305), the National Fundamental Research
Program (Grant No. 2001CB309300), and National Science
Fund for Distinguished Young Scholars (Grant No.
10425524).


\begin{references}
\bibitem{note2} Note that in order to reach the above results, we need to apply an additional local unitary transformation to the resulted state after
each measurement.
\bibitem{note3} In deriving all formulae in this paper, we assume that $p$ is much smaller than $1$ for simplicity
of the calculations. However, it is worth noting that our scheme works effectively
when $p$ is comparable with $0.5$ or even larger.
\end{references}
\end{document}